\documentclass[pra, twocolumn,amsfonts,amsmath,amssymb,eufrak]{revtex4-1}
\usepackage{epsfig}
\usepackage{color}
\usepackage{bm}
\usepackage{hyperref}% add hypertext capabilities
\usepackage[dvipsnames]{xcolor}
\hypersetup{
    bookmarksnumbered=true, % If Acrobat bookmarks are requested, include section numbers
    unicode=false, % non-Latin characters in Acrobat bookmarks
    pdfstartview={FitH}, % fits the width of the page to the window
    pdftitle={}, % title
    pdfauthor={}, % author
    pdfsubject={}, % subject of the document
    pdfcreator={}, % creator of the document
    pdfproducer={}, % producer of the document
    pdfkeywords={}, % list of keywords
    pdfnewwindow=true, % links in new window
    colorlinks=true, % false: boxed links; true: colored links
    linkcolor=NavyBlue, % color of internal links
    citecolor=NavyBlue, % color of links to bibliography
    filecolor=NavyBlue, % color of file links
    urlcolor=NavyBlue % color of external links
}
\usepackage{braket}

\usepackage{graphicx}
\usepackage{amsthm}
\usepackage{bookmark}

\newcommand{\eq}[1]{\begin{equation} #1 \end{equation}}
\newcommand{\eqa}[2]{\begin{equation} #1 \label{#2} \end{equation}}
\newcommand{\balign}[1]{\begin{align} #1 \end{align}}

\newcommand{\mx}[1]{\begin{pmatrix}#1 \end{pmatrix}}
\newcommand{\ul}{\underline}

\newcommand{\figin}[4]
{\begin{figure}[tb]
\includegraphics[width= #1]{#2.pdf}
\caption{#3}
\label{f:#4}
\end{figure}}

\newcommand{\todayd}{\the\year/\the\month/\the\day}

\newcommand{\bib}{\bibitem}

\newcommand{\lr}{\leftrightarrow}

\newcommand{\lmd}{\lambda}

\newcommand{\lb}{\label}
\newcommand{\nt}{\notag}
\newcommand{\Tr}{\mathrm{Tr}}
\newcommand{\ft}[2]{\left. #1 \right|_{#2}}

\newcommand{\bel}{\begin{easylist}}
\newcommand{\eel}{\end{easylist}}
\newcommand{\bi}[1]{\begin{itemize} #1 \end{itemize}}

\newcommand{\eref}[1]{Eq.~\eqref{#1}}

\newcommand{\lref}[1]{Lemma.~\ref{t:#1}}

\newcommand{\fref}[1]{Fig.~\ref{f:#1}}
\newcommand{\sref}[1]{Sec.~\ref{s:#1}}

\def \({\left(}
\def \){\right)}
\def \[{\left[}
\def \]{\right]}

\newcommand{\abs}[1]{\left|#1\right|}

\newcommand{\sumtwo}[2]%
{\mathop{\sum_{#1}}_{#2}}
\newcommand{\sumthree}[3]%
{\mathop{\mathop{\sum_{#1}}_{#2}}_{#3}}
\newcommand{\sumfour}[4]%
{\mathop{\mathop{\mathop{\sum_{#1}}_{#2}}_{#3}}_{#4}} 
\newcommand{\prodtwo}[2]%
{\mathop{\prod_{#1}}_{#2}}
\newcommand{\mintwo}[2]%
{\mathop{\min_{#1}}_{#2}}
\newcommand{\maxtwo}[2]%
{\mathop{\max_{#1}}_{#2}}
\newcommand{\maxthree}[3]%
{\mathop{\mathop{\max_{#1}}_{#2}}_{#3}}
\newcommand{\limtwo}[2]%
{\mathop{\lim_{#1}}_{#2}}
\newcommand{\suptwo}[2]%
{\mathop{\sup_{#1}}_{#2}}
\newcommand{\supthree}[3]%
{\mathop{\mathop{\sup_{#1}}_{#2}}_{#3}}
\newcommand{\supfour}[4]%
{\mathop{\mathop{\mathop{\sup_{#1}}_{#2}}_{#3}}_{#4}} 
\newcommand{\inftwo}[2]%
{\mathop{\inf_{#1}}_{#2}}
\newcommand{\infthree}[3]%
{\mathop{\mathop{\inf_{#1}}_{#2}}_{#3}}
\newcommand{\inffour}[4]%
{\mathop{\mathop{\mathop{\inf_{#1}}_{#2}}_{#3}}_{#4}} 
\newcommand{\bsA}{\boldsymbol{A}}
\newcommand{\bsB}{\boldsymbol{B}}

\newcommand{\bsS}{\boldsymbol{S}}

\newcommand{\bbR}{\mathbb{R}}

\newcommand{\Di}{\mathit{\Delta}}

\newcommand{\para}[1]{{\em #1}\/.---}

\newtheorem{thm}{Theorem}
\newtheorem{lm}[thm]{Lemma}
\newtheorem{pro}[thm]{Proposition}

\newtheorem*{thmn}{Theorem}

\newcommand{\bpf}[1]{\begin{proof} #1 \end{proof}}

\newcommand{\blm}[1]{\begin{lm} #1 \end{lm}}

\theoremstyle{definition}
\newtheorem{dfn}{Definition}

\newcommand{\ba}[2]{
\begin{array}{#1}
#2
\end{array}
}

\def\rnum#1{\resizebox{0.5em}{\height}{\expandafter{\romannumeral #1}}}
\def\Rnum#1{\resizebox{0.5em}{\height}{\uppercase\expandafter{\romannumeral #1}}}
\begin{document}

% \preprint{APS/123-QED}

\newcommand{\nameoftitle}{Dichotomy theorem separating complete integrability and non-integrability of isotropic spin chains}

\title{
\nameoftitle
}% Force line breaks with \\

\author{Naoto Shiraishi} 
\email{shiraishi@phys.c.u-tokyo.ac.jp}
\affiliation{Graduate school of arts and sciences, The University of Tokyo, 3-8-1 Komaba, Meguro-ku, Tokyo 153-8902, Japan}%

\author{Mizuki Yamaguchi} 
\email{yamaguchi-q@g.ecc.u-tokyo.ac.jp }
\affiliation{Graduate school of arts and sciences, The University of Tokyo, 3-8-1 Komaba, Meguro-ku, Tokyo 153-8902, Japan}%

%\date{}% It is always \today, today,
             %  but any date may be explicitly specified

\begin{abstract}
We investigate the integrability and non-integrability of isotropic spin chains with nearest-neighbor interaction with general spin $S$ in terms of the presence or absence of local conserved quantities.
We prove a dichotomy theorem that whether a single quantity is zero or not sharply separates two scenarios: (i) this system has $k$-local conserved quantities for all $k$ (completely integrable), or (ii) this system has no nontrivial local conserved quantity (non-integrable).
This result excludes the possibility of an intermediate system with some but not all local conserved quantities, which solves in the affirmative the Grabowski-Mathieu conjecture.
This theorem also serves as a complete classification of integrability and non-integrability for $S\leq 13.5$, suggesting that all the integrable models are in the scope of the Yang-Baxter equation.
\end{abstract}

\maketitle

\twocolumngrid

\para{Introduction}
Integrability and non-integrability of quantum many-body systems are important subjects in theoretical physics.
Integrable systems have been studied for almost 100 years, and various quantum systems such as the Heisenberg model, the transverse Ising model, and more complicated models were discovered as integrable models~\cite{JM, Bax}.
We note that unlike classical integrability there is no universally-accepted definition of quantum integrability and non-integrability, though known integrable models with local interactions share common properties including solvability and infinitely many local conserved quantities.
Following previous studies~\cite{CM11, GE16, Shi19}, in this Letter, we characterize complete integrability and non-integrability as having all $k$-local conserved quantities and no nontrivial local conserved quantities, respectively.

To find integrable models, a sophisticated technique based on the Yang-Baxter equation, named the algebraic Bethe ansatz or the quantum inverse scattering methods, works as a powerful tool~\cite{Fad, KBI}.
Precisely, if the Yang-Baxter equation is satisfied, then we can construct an infinite series of local conserved quantities~\cite{DG, FF, GM94, GM95-1, NF20}, with which energy eigenstates can also be constructed.
In this case, we also have a boost operator, which generates local conserved quantities through a simple calculation~\cite{GM94, GM95-1}.
It is noteworthy, however, that some important integrable models including the Hubbard model and the XY model also have an infinite family of local conserved quantities but neither satisfy the conventional (differentiable) Yang-Baxter equation nor have a boost operator~\cite{GM95-2, Lin01, Lee21}.

In contrast to integrability, theoretical aspects of non-integrability have not yet been fully addressed, though the absence of nontrivial local conserved quantities is a relevant property to various phenomena including thermalization~\cite{Lan, EF, SM17, MS17, SMreply} and response to external drives~\cite{Kub, SK, SF, Zot}.
%One reason for this insufficiency is that the proof of non-existence (no solution or no nontrivial local conserved quantity) is usually harder than the proof of existence (of solutions and nontrivial local conserved quantities).
%Despite the scarcity of studies focusing on non-integrability, the existing literature can broadly be classified into two approaches.
Within the sparse literature devoted to non-integrability, one traditional approach is restricting the solution method to the Yang-Baxter equation and testing the validity of its lowest-order relation (the Reshetikhin condition)~\cite{KS82, Ken92, BY94, BN14, Hie92, LPR19, Lee20, Lee21, Mai24, LP24}.
%With this approach, several classes of spin chains have been classified into solvable and unsolvable ones in terms of the Yang-Baxter equation.
%As far as previous research has explored, all models satisfying the lowest-order Yang-Baxter equation (the Reshetikhin condition~\cite{KS82, Ken92}) also satisfy the full-order Yang-Baxter equation, implying the solvability.
However, this approach suffers from an inherent limitation: we cannot rule out the possibility that a model is not solvable by the Yang-Baxter equation but is still solvable by another, possibly yet unknown, solution method.
A more recent approach directly proves the absence of nontrivial local conserved quantities, which was invented by Ref~\cite{Shi19}, and by applying this technique various quantum spin systems have been proven to be non-integrable~\cite{Chi24-1, Shi24, PL24-1, CY24, ST24, Chi24-2, PL24-2, HYC24, YCS24-1, YCS24-2, Shi25, FT25}.
Notably, symmetric $S=1/2$ systems in one~\cite{YCS24-1, YCS24-2, Shi25} and higher dimensions~\cite{ST24, Chi24-2} are completely classified into completely integrable and non-integrable.
Another important progress is brought by Hokkyo~\cite{Hok25}, which observes similarity among many previous non-integrability proofs and clarifies the sufficient conditions for proving non-integrability.
Precisely, it is shown that there are two key operator equations and if these two operator equations respectively have specific forms of solutions as unique solutions, then the non-integrability proof surely works.

In principle, there may exist an intermediate system which has some $k$-local conserved quantities but not all ones.
On the other hand, all known models have (i) all $k$-local conserved quantities, or (ii) no nontrivial local conserved quantity, and we have never found such intermediate systems.
Therefore, we strongly expect the dichotomy that only the above two extremes are possible.
A similar conjecture was carried out by Grabowski and Mathieu~\cite{GM95-2}, observing that the presence or absence of a 3-local conserved quantity determines the complete integrability or non-integability.
%The Grabowski-Mathieu conjecture and its refinement~\cite{GP21} are expected to serve as the guidepost of investigation of unexplored systems.
However, proving this dichotomy on a rigorous foundation has been left as an open problem.

In this Letter, we rigorously establish this dichotomy for isotropic spin chains with general spin $S$.
Precisely, we prove that whether a single quantity $D^3$ is zero or not fully determines whether this model has all $k$-local conserved quantities (completely integrable) or no nontrivial local conserved quantity (non-integrable).
This confirms the all-or-nothing principle of local conserved quantities, excluding the possibility of an intermediate system with some local conserved quantities.
The separation into the two extreme scenarios applies all isotropic spin chains including unexplored ones, which indeed meets the aim of the Grabowski-Mathieu conjecture.
Notably, our criterion of this dichotomy is easy to compute, which shows clear contrast to existing attempts~\cite{GM95-2, Hok25} requiring to solve complex search problems of operator equations.

\para{Setup and main result}
We consider a shift-invariant isotropic spin chain with general spin $S$ with nearest-neighbor interaction.
The isotropic condition is equivalent to the $SU(2)$ invariance.
The system Hamiltonian in consideration is expressed as
\balign{
H=&\sum_i h_{i,i+1}, \hspace{15pt} h_{i,i+1}=\sum_{n=1}^{2S} J_n (\bsS_i \cdot \bsS_{i+1})^n. \lb{H}
}
We here introduce a key quantity
\eqa{
D=[[h_{i,i+1}, h_{i+1,i+2}], h_{i,i+1}+ h_{i+1,i+2}]
}{D-def}
and denote its 3-support operator terms by $D^{3}$.

\figin{8.5cm}{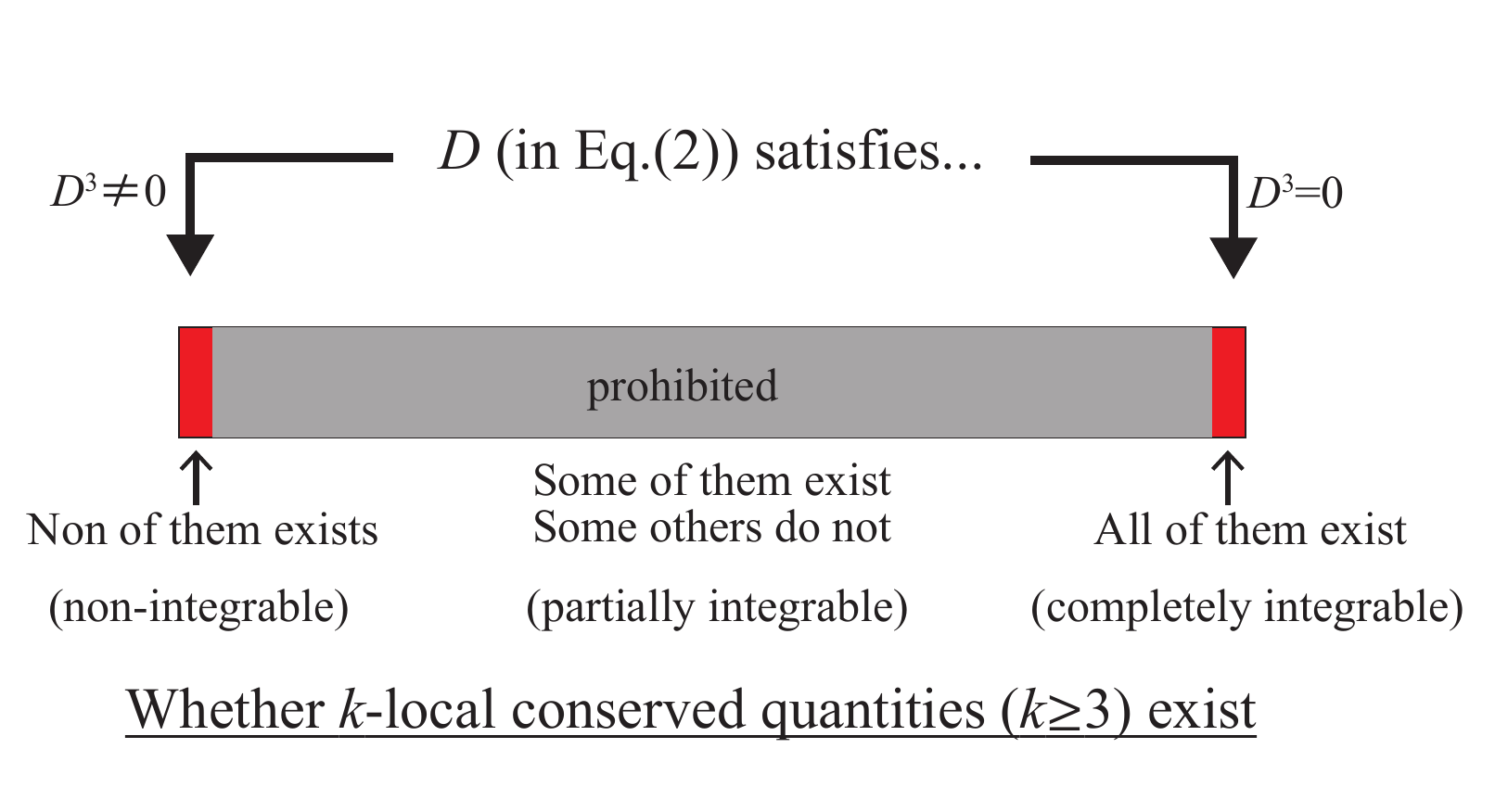}{
Claim of our dichotomy theorem.
If $D^3=0$, then this model has all $k$-local conserved quantities (completely integrable).
If $D^3\neq 0$, then this model has no nontrivial local conserved quantity (non-integrable).
}{flow}

Now we state the dichotomy theorem on the isotropic spin chains (see also \fref{flow}).
\begin{thmn}
Consider a shift-invariant isotropic spin chain with general spin $S$ with nearest-neighbor interaction, whose Hamiltonian is given by \eref{H}.
For the above class of spin chains, 
\bi{
\item If $D^3=0$, then this model has $k$-local conserved quantities for all positive integers $k\geq 3$ (completely integrable)
\item If $D^3\neq 0$, then this model has no nontrivial local conserved quantity (non-integrable)
}
\end{thmn}

Here, in the latter case we exclude all $k$-local conserved quantities with $3\leq k\leq N/2-1$.
Our result establishes the dichotomy of local conserved quantities (all or nothing), excluding a {\it partially-integrable system}~\cite{YCS24-1, YCS24-2} with a finite number of nontrivial local conserved quantities.
In addition, the criterion of this dichotomy is whether a single quantity is zero or not, which is easy to calculate.
This shows a high contrast to the Grabowski-Mathieu conjecture~\cite{GM95-2, GP21} and its partial proof by Hokkyo~\cite{Hok25}, where we need to solve search problems on operator equations to confirm non-integrability.

Note that in the case of $D^3=0$ the $k$-local conserved quantities $Q^k$ is generated recursively by the boost operator in the form of $B=\sum_j j h_{j,j+1}$ as
\eqa{
Q^{k+1}=[B, Q^k]
}{Qk-recursive}
with $Q^2=H$.
The 3-local conserved quantity, for example, is given by
\balign{
Q^3=\sum_i &[h_{i,i+1}, h_{i+1,i+2}], \lb{Q3} %\\
%Q^4=\sum_i &2[[h_{i,i+1}, h_{i+1,i+2}], h_{i+2,i+3}] \nt \\
%&+[[h_{i,i+1}, h_{i+1,i+2}], h_{i+1,i+2}]. \lb{Q4}
}
%This boost operator is conjectured to generate higher conserved quantities $Q^n$ for $n\geq 5$ (conjecture 2 of Ref.~\cite{GM95-2}), while whether the generated operators are indeed conserved quantities is an open problem.
with which $[Q^3,H]$ is equivalent to the shift sum of $D$ given in \eref{D-def}.

\para{Yang-Baxter equation and Reshetikhin condition}
%Importantly, the conservation of the 3-local conserved quantity, $[Q^3,H]=0$, is equivalent to the well-known Reshetikhin condition~\cite{KS82, Ken92}, which is equivalent to the lowest-order relation in the Yang-Baxter equation.
In order to explain the significance of our result, we here briefly introduce the Yang-Baxter equation and the Reshetikhin condition (for details, see the Supplemental Material~\cite{SM}).
The Yang-Baxter equation is a relation on a one-parameter family of a matrix $R_{ij}(\lmd)$ on two sites $i$ and $j$ given by
\eqa{
R_{12}(\lmd)R_{23}(\lmd+\mu)R_{23}(\mu)=R_{23}(\mu)R_{13}(\lmd+\mu)R_{12}(\lmd).
}{YB}
The $R$ matrix can be expanded as
\eqa{
R(\lmd)=\Pi+\lmd \Pi h +\lmd^2 \frac12 \Pi h^2+\cdots,
}{R-expand}
where $h$ is the local Hamiltonian and $\Pi$ is a swap operator.
The third and higher-order terms are not universal.
Plugging \eref{R-expand} into the Yang-Baxter equation \eqref{YB} and compare the lowest nontrivial order, $\lmd\mu^2$, we obtain the {\it Reshetikhin condition} claiming that $D$ in \eref{D-def} can be expressed as~\cite{KS82, Ken92}
\eqa{
D_{i,i+1,i+2}=E_{i+1,i+2}-E_{i,i+1},
}{Res}
where $E_{ij}$ is a proper operator on two sites $i$ and $j$.

%Thus, the Reshetikhin condition is the nontrivial lowest-order relation of the Yang-Baxter equation.
%The Reshetikhin condition is also equivalent to the relation $[Q^3,H]=0$ with $Q^3$ given by \eref{Q3}, meaning the presence of 3-local conserved quantity in this form.
%Note that the Reshetikhin condition also implies $[Q^4, H]=0$ with $Q^4$ given by \eref{Q4} (see Supplemental Material~\cite{SM}).

\para{Implications of the main theorem}
Remarkably, our theorem confirms that the Reshetikhin condition does not miss integrable systems in isotropic spin chains, justifying the approach to non-integrability based on the Reshetikhin condition explained in the introduction (i.e., asserting non-integrability by the violation of the Reshetikhin condition)~\cite{Ken92, BY94}.
The integrable part ($D^3=0$) of our theorem contains the Reshetikhin condition, since the Reshetikhin condition~\eqref{Res} is equivalent to the condition $[Q^3,H]=0$ with \eref{Q3}.
Note that our dichotomy theorem imposes that if an infinite family of local conserved quantities exists, the Reshetikhin condition should be satisfied.
Hence, our theorem establishes that the Reshetikhin condition is the necessary and sufficient condition for complete integrability in isotropic spin chains. 
In light of the fact that some integrable models including the Hubbard model do not satisfy the Reshetikhin condition, it is striking that all isotropic integrable models admit the Reshetikhin condition.

%We briefly explain this point later, and present the details of the boost operator, the Reshetikhin condition, and the Yang-Baxter equation in \cite{SM}.

% the lowest-order Yang-Baxter equation (the Reshetikhin condition) and

%First, this theorem confirms that all integrable isotropic spin chains should have 3-local and 4-local conserved quantities in the form of Eqs.~\eqref{Q3} and \eqref{Q4}.
%Our theorem excludes the existence of (unknown) integrable systems without 3-local and 4-local conserved quantities.
In addition, our theorem proves the Grabowski-Mathieu conjecture (both conjecture 1 and 2)~\cite{GM95-2, GP21} in the affirmative in the class of isotropic spin chains, stating that a system is integrable if and only if it has a 3-local conserved quantity.
This is the first proof of the Grabowski-Mathieu conjecture for a large class of spin systems including models which we have not yet explored.

%This approach is applied to more general spin systems~\cite{BN14, Hie92, LPR19, Lee20, Mai24, LP24}, which is expected also to be justified by similar dichotomy proofs.

This theorem also serves as the complete classification of complete integrability and non-integrability up to $S=13.5$.
Kennedy~\cite{Ken92} and Batchelor and Yung~\cite{BY94} have examined all isotropic spin chains satisfying the Reshetikhin condition less than or equal to $S=13.5$ by computer search and found that all the spin chains satisfying the Reshetikhin condition are in fact exhausted by known integrable systems solvable by the Yang-Baxter equation.
Precisely, there are four infinite sequences of integrable systems and one exceptional integrable system for $S=3$ (see End matter), which correspond to four infinite series of compact Lie groups and one exceptional Lie group satisfying SU(2) symmetry~\cite{BY94}.
Combining this fact and our theorem, we conclude that these four infinite sequences of integrable systems and one exceptional integrable system are indeed the complete list of integrable isotropic spin chains up to $S=13.5$.

\para{Useful basis}
To prove our main theorem, we employ a useful basis of operators on spin $S$ named {\it noncommutative spherical harmonics}, which is also known as {\it fuzzy harmonics}~\cite{Iso01, LV14, AA19}.
The basis operator $Y^{lm}$ has two indices, the azimuthal quantum number $l$ and the magnetic quantum number $m$, which take integers satisfying $0\leq l\leq 2S$ and $-l\leq m\leq l$.
The product of two noncommutative spherical harmonics is given by
\eqa{
Y^{l_1,m_1}Y^{l_2,m_2}=\sum_l K^{l_1,l_2,l}_{m_1,m_2}Y^{l, m_1+m_2},
}{YYK}
where $K^{l_1,l_2,l}_{m_1,m_2}$ is a coefficient proportional to the well-known Clebsch–Gordan coefficient.
For our purpose, important properties of $K$ are the following two:
\bi{
\item $K^{l_1,l_2,l}_{m_1,m_2}$ is zero if the triangular condition $\abs{l_1-l_2}\leq l\leq l_1+l_2$ is not satisfied.
\item $K^{l_1,l_2,l}_{m_1,m_2}=K^{l_2,l_1,l}_{m_2,m_1}$ if $l_1+l_2+l$ is even, and $K^{l_1,l_2,l}_{m_1,m_2}=-K^{l_2,l_1,l}_{m_2,m_1}$ if $l_1+l_2+l$ is odd.
}
The latter fact implies that a commutator $[Y^{l_1,m_1}, Y^{l_2,m_2}]$ consists of $Y^{l', m_1+m_2}$ such that $l'$ and $l_1+l_2$ are different parities, and an anti-commutator $\{ Y^{l_1,m_1}, Y^{l_2,m_2}\}$ consists of $Y^{l', m_1+m_2}$ such that $l'$ and $l_1+l_2$ are the same parity.
Using this basis, the local Hamiltonian $h$ given in \eref{H} is also expanded as~\cite{SM}
\eqa{
h_{i,i+1}=\sum_{l=0}^{2S} c_l \( \sum_{m=-l}^l Y^{l,m}_iY^{l,-m}_{i+1}\) ,
}{YYform}
where we renormalized $Y^{lm}$ such that the coefficients of $Y^{l,m}_iY^{l,-m}_{i+1}$ to be one in this expression.

\para{Proof for $D^3=0$}
We here consider the case of $D^3=0$.
We here employ a recent result~\cite{Hok25-2} stating that if the 3-local conserved quantity $Q^3$ given by \eref{Q3} conserves ($[Q^3,H]=0$), then all $k$-local conserved quantities $Q^k$ given by \eref{Qk-recursive} also conserve ($[Q^k,H]=0$) (see \lref{Q3-Qm} in the Supplemental Material~\cite{SM}).
Hence, it suffices to show the presence of a 3-local conserved quantity, or the Reshetikhin condition.
By comparing the condition $D^3=0$ to the Reshetikhin condition \eqref{Res}, our task boils down to that 2-support operators in $D$ take the form of $E_{i+1,i+2}-E_{i,i+1}$.
We shall show this fact as follows.

By assumption, $[[h_{i,i+1}, h_{i+1,i+2}], h_{i+1,i+2}]$ is an (at most) 2-support operator, and since site $i$ has a nontrivial action, this 2-support operator is on $i$ and $i+1$, which we denote by
\eq{
[[h_{i,i+1}, h_{i+1,i+2}], h_{i+1,i+2}]=A_{i,i+1}.
}
Considering its reflection, we have
\balign{
[[h_{i,i+1}, h_{i+1,i+2}], h_{i, i+1}]=&-[[h_{i+1,i+2}, h_{i,i+1}], h_{i, i+1}] \nt \\
=&-A_{i+2,i+1}.
}
Thus, it suffices to show that $A$ has inversion symmetry;
\eq{
A_{ij}=A_{ji},
}
under which we can set $A_{ij}=E_{ij}$.
We can show the inversion symmetry as follows.
Since both $[h_{i,i+1}, h_{i+1,i+2}]$ and $h_{i+1,i+2}$ are $SU(2)$ symmetric, their commutator $[[h_{i,i+1}, h_{i+1,i+2}], h_{i+1,i+2}]$ should also be $SU(2)$ symmetric.
Notice that all 2-support $SU(2)$ symmetric operator takes the form of the right-hand side of \eref{H}, which also has inversion symmetry.

\para{Proof for $D^3\neq 0$}
We next consider the case of $D^3\neq 0$.
We prove the non-integrability of this Hamiltonian with the help of the scheme proposed by Hokkyo~\cite{Hok25}.
Below we denote the $m$-support part of $k$-local conserved quantity by $Q^{k,m}$, i.e., we expand $Q^k$ as
\eq{
Q^k=\sum_{m=1}^k Q^{k,m}.
}

We first restrict the possible form of the $k$-support operator $Q^{k,k}$.
We here introduce the {\it doubling-product operators} $\Di^k$ which is recursively defined as
\eq{
\Di^2:=h, \hspace{10pt} \Di^k_{1,2,\ldots ,k}:=[\Di^{k-1}_{1,2,\ldots , k-1}, h_{k-1,k}].
}
It is shown in Ref.~\cite{Hok25} that $Q^{k,k}$ should be proportional to $\Di^k$ if $f(A)=[I_1\otimes A_2, h_{12}]$ is injective and the equation for 2-support operator $X$
\eqa{
[X_{i,i+1}, h_{i+1,i+2}]+[X_{i+1,i+2}, h_{i,i+1}]=0
}{step1-cond}
admits only solutions for $X$ proportional to $h$.
We provide the proof outline of satisfying these two conditions below.
Their full proofs are shown in the Supplemental material~\cite{SM}.

The former injective condition is confirmed as follows:
Using the projection operator formula in the character theory, we show that if $f(A)=0$, then $f(A^l)=0$ for all $l$ where $A^l$ is the projection of $A$ onto the space with the azimuthal quantum number $l$.
Note that $A^l$ is expanded as $A^l=\sum_m k_m Y^{l,m}$ with coefficient $k_m$.
However, for any $l$, $m$ and $l'$, there exists $m'$ such that $[Y^{l,m}, Y^{l',m'}]\neq 0$.
Comparing \eref{YYform}, we conclude that $f(A^l)=0$ happens only if $A^l$ is trivial (i.e., proportional to the identity operator).

We next treat the latter condition.
We expand $X$ as
\eq{
X=\sum_{l,l',m,m'}q_{l,m,l',m'}Y^{l,m}_iY^{l',m'}_{i+1}
}
and specify coefficients $q_{l,m,l',m'}$ step by step.
Let $L:=\{l|c_l\neq 0\}$ be the set of integers $l$ with nonzero coefficient $c_l$ in \eref{YYform}.
It is easy to show that a term $Y^{l,m}Y^{l',m'}$ appears in $X$ only if $l,l'\in L$.
A key observation is that $Y^{l,m}Y^{l',m'}$ with a nonzero coefficient should take the form of $Y^{l,m}Y^{l,-m}$.
To prove this, we employ a lemma that the coefficient of $Y^{l,m}Y^{l',m'}$ and that of $Y^{l,m\pm 1}Y^{l',m'\mp 1}$ are linearly connected:
\eq{
q_{l,m,l',m'}\propto q_{l,m\pm 1, l', m'\mp1}.
}
Applying this relation repeatedly, only one of $m$ or $m'$ finally exceeds $l$ or $l'$ if $(l',m')\neq (l,-m)$, which implies zero coefficient.
Since all the coefficients in this sequence are linearly connected, we arrive at $q_{l,m,l',m'}=0$ if $(l',m')\neq (l,-m)$.
We finally confirm that all the coefficients of $Y^{l,m}Y^{l,-m}$ are proportional to those of $h$, which is the desired relation.
%In particular, we have
%\eqa{Q^{3,3}=[h_{i,i+1},h_{i+1,i+2}].}{Q33}

%We next apply the Hokkyo scheme for step 2.
We then show that $Q^{k,k}$ is in fact zero.
A key fact to show this claim is that linear relations in $[Q^k,H]=0$ for $k$-support operators can be reduced to those in $[Q^3,H]=0$ for 3-support operators~\cite{Hok25}.
In other words, if we show $Q^{3,3}=0$ only from the relation that all coefficients of 3-support operators in
\balign{
&[Q^{3,3}_{i,i+1,i+2}, h_{i,i+1}]+[Q^{3,3}_{i,i+1,i+2}, h_{i+1,i+2}] \nt \\ &
+[Q^{3,2}_{i,i+1}, h_{i+1,i+2}]+[Q^{3,2}_{i+1,i+2}, h_{i,i+1}]
}
are zero, then we have proof of the absence of $k$-local conserved quantities with general $k$.
By comparing the condition of our main theorem and the above relation, it suffices to show $Q^{3,2}=0$.
Below, we show that this relation is a consequence of the $SU(2)$ symmetry and the time-reversal symmetry of the Hamiltonian \eqref{H}.

First, since both $H$ and $Q^{3,3}$ have $SU(2)$ symmetry, the entire conserved quantity $Q^3$, and thus $Q^{3,2}$, should also have $SU(2)$ symmetry.
An $SU(2)$-symmetric 2-support operator is expressed as in the form of the right-hand-side of \eref{H}, or equivalently \eref{YYform}.

On the other hand, since $H$ has time-reversal symmetry, the conserved quantity $Q^3$ is either time-symmetric or time-antisymmetric.
These two cases correspond to the cases that the sum of all azimuthal quantum numbers in the operator is even (time-symmetric) and odd (time-antisymmetric).
Since $Q^{3,3}$ is equal to the doubling-product operator $\Di^3$, which is time-antisymmetric, $Q^3$, and thus $Q^{3,2}$, should also be time-antisymmetric.
However, any $SU(2)$-symmetric 2-support operator (the right-hand-side of \eref{YYform}) is time-symmetric, and thus $Q^{3,2}$ should be zero, which completes the proof.

\para{Discussion}
We have proven that an isotropic spin chain has (i) all $k$-local conserved quantities (completely integrable), or (ii) no nontrivial local conserved quantity (non-integrable).
We have also shown that whether a single quantity $D^3$ given in \eref{D-def} is zero or not sharply distinguishes these two possible scenarios.
Our result solves the Grabowski-Mathieu conjecture~\cite{GM95-2, GP21} in the affirmative for isotropic spin chains.
Our result also serves as the classification theorem of integrability and non-integrability for the case with less than or equal to $S=13.5$.

In our theorem, a single quantity $D^3$ serves as a criterion to distinguish the complete integrability and non-integrability, which is an advantage compared to previously proposed criteria~\cite{GM95-2, GP21, Hok25} containing search problems of operator equations.
%Provided conjecture 2 by Grabowski and Mathieu~\cite{GM95-2}, the integrability or non-integrability of isotropic spin chains is completely determined by the same single relation.
One may expect an extension of the same criterion to a wider class of systems beyond isotropic ones.
However, the Hubbard model, which has many symmetries close to isotropy, has all $k$-local conserved quantity but $D^3\neq 0$.
This reflects the fact that the local conserved quantities of the Hubbard model are generated not by the standard boost operator $B$ but its generalized one~\cite{Lin01, Lee21}.
In spite of this negative suggestion, it is still plausible that a simple criterion similar to $D^3$ determines integrability and non-integrability for a wider class of spin systems beyond isotropic spin chains, which is worth investigating.

Our result shows that all integrable isotropic spin chains satisfy the lowest-order Yang-Baxter equation (the Reshetikhin condition).
Moreover, restricting isotropic spin chains with $S\leq 13.5$, all integrable isotropic spin chains satisfy the Yang-Baxter equation, meaning that the Yang-Baxter equation is the unique solution method in this regime.
%This shows high contrast to the Hubbard model, which is solvable but does neither satisfy the lowest-order Yang-Baxter equation nor have a boost operator generating local conserved quantities.
On the other hand, again the Hubbard model does not admit the original Yang-Baxter equation~\cite{Lee21}.
On the basis of the above, we expect that the Yang-Baxter equation is not always universal but indeed universal in some classes including isotropic spin chains.
We note that the technique of non-integrability proofs is also extended to write down local conserved quantities of integrable systems~\cite{NF20, YF23}.
The non-integrability proof approach will pave the way to clarify how universally the Yang-Baxter equation applies, which merits further research.

%\begin{acknowledgments}
{\it Acknowledgments.} ---
It is a pleasure to thank Yuki Furukawa, for his essential contribution to \lref{l-decompose} on the character theory.
We also thank Akihiro Hokkyo, Yuuya Chiba, and Hal Tasaki for the fruitful discussion.
This work is supported by JST ERATO Grant No. JPMJER2302, Japan.

%\end{acknowledgments}

\let\oldaddcontentsline\addcontentsline% Store \addcontentsline
\renewcommand{\addcontentsline}[3]{}% Make \addcontentsline a no-op

\section*{End matter}

\subsection*{List of known isotropic integrable models solvable through the Yang-Baxter equation}

Following Refs.~\cite{Ken92, BY94}, we here list the known integrable isotropic Hamiltonians solvable by the Yang-Baxter equation.
Below we denote by $P^{(a)}$ the projection operator acting on two sites projecting onto the subspace with total spin $S=a$.
Using this projection operator and the swap operator $\Pi$, the local Hamiltonians of four infinite series for spin $S$ are given by
\balign{
h^{\rm I}=&\Pi, \\
h^{\rm II}=&\( S+\frac12-(-1)^{2S}\) \Pi-(-1)^{2S}(2S+1)P^{(0)}, \\
h^{\rm III}=&\sum_{a=0}^{2S}\( \sum_{j=1}^a \frac 1j\) P^{(a)}, \\
h^{\rm IV}=&P^{(0)},
\intertext{
and one exceptional local Hamiltonian for $S=3$ is given by
}
h^{\rm V}=&11P^{(0)}+7P^{(1)}-17P^{(2)}-11P^{(3)} \nt \\
&-17P^{(4)}+7P^{(5)}-17P^{(6)}.
}
We note that the projection operator and the swap operator are polynomials of $T:=\bsS_i\cdot \bsS_{i+1}$ as
\balign{
P^{(a)}=&\prod_{n=0 \ (n\neq a)}^{2S}\frac{T-x_n}{x_a-x_n} \\
\Pi=&(-1)^{2S}\sum_{a=0}^{2S}(-1)^a P^{(a)}
}
with $x_n=\frac12 n(n+1)-S(S+1)$, which confirms that these Hamiltonians are indeed isotropic.

\subsection*{Formal definitions of noncommutative spherical harmonics and coefficient $K$}

The matrix elements of the noncommutative spherical harmonics for spin $S$ before renormalization (in \eref{YYform}) in the eigenbasis of $S^z$ is given by~\cite{LV14}
\eq{
\braket{i|Y^{l,m}|j}=\sqrt{\frac{4S+1}{2l+1}}(-1)^{2S-l+m} C^{l,-m}_{2S,i,2S,-j}
}
with $-2S\leq i,j\leq 2S$, where $C$ is the Clebsh-Gordan coefficient.
Detailed properties of $Y^{l,m}$ are presented in Refs.~\cite{LV14, AA19}.

The coefficient $K^{l_1,l_2,l}_{m_1,m_2}$ is defined by using the 6j-symbol as~\cite{AA19}
\balign{
K^{l_1,l_2,l}_{m_1,m_2}:=&(-1)^{4S+l}\sqrt{\frac{(2l_1+1)(2l_2+1)(4S+1)}{4\pi}} \nt \\
&\cdot \mx{l_1&l_2&l\\ 2S&2S&2S}C^{l,m_1+m_2}_{l_1,m_1,l_2,m_2}.
}

\subsection*{Examples of noncommutative spherical harmonics for $S=1/2$ and $S=1$}

We here present the noncommutative spherical harmonics for $S=1/2$ and $S=1$ up to a constant in the matrix form.
For $S=1/2$, they are given by
\balign{
Y^{0,0}&=\mx{1&0 \\ 0&1}, \hspace{10pt}
Y^{1,0}=\mx{1&0 \\ 0&-1}, \nt \\
Y^{1,1}&=\mx{0&1\\ 0&0}, \hspace{10pt}
Y^{1,-1}=\mx{0&0 \\ 1&0}.
}
For $S=1$, they are given by
\balign{
Y^{0,0}&=\mx{1&0&0 \\ 0&1&0 \\ 0&0&1}, \hspace{10pt}
Y^{1,0}=\mx{1&0&0 \\ 0&1&0 \\ 0&0&-1}, \nt \\
Y^{1,1}&=\mx{0&1&0 \\ 0&0&1 \\ 0&0&0}, \hspace{10pt}
Y^{1,-1}=\mx{0&0&0 \\ 1&0&0 \\ 0&1&0}, \nt \\
Y^{2,0}&=\mx{1&0&0 \\ 0&-2&0 \\ 0&0&1}, \hspace{10pt}
Y^{2,1}=\mx{0&1&0 \\ 0&0&-1 \\ 0&0&0}, \nt \\
Y^{2,-1}&=\mx{0&0&0 \\ 1&0&0 \\ 0&-1&0}, \hspace{10pt}
Y^{2,2}=\mx{0&0&1 \\ 0&0&0 \\ 0&0&0}, \nt \\
Y^{2,-2}&=\mx{0&0&0 \\ 0&0&0 \\ 1&0&0}.
}

\let\addcontentsline\oldaddcontentsline% Restore \addcontentsline

\clearpage

\pagestyle{plain}

\setcounter{page}{1}
\setcounter{figure}{0}
\renewcommand{\thefigure}{S.\arabic{figure}}

%%%%%%%%%%%%%%%%%%%%%%%%%%%%%%%%%%%%%%%%%%%
%%%%%%%%%%%%%%%%%%%%%%%%%%%%%%%%%%%%%%
% To modify figure captions
\makeatletter
\long\def\@makecaption#1#2{{
\advance\leftskip1cm
\advance\rightskip1cm
\vskip\abovecaptionskip
\sbox\@tempboxa{#1: #2}%
\ifdim \wd\@tempboxa >\hsize
 #1: #2\par
\else
\global \@minipagefalse
\hb@xt@\hsize{\hfil\box\@tempboxa\hfil}%
\fi
\vskip\belowcaptionskip}}
\makeatother
%%%%%%%%%%%%%%%%%%%%%%%%%%%%%%%%%%%%%%
\newcommand{\vo}{\upsilon}
\newcommand{\midskip}{\vspace{3pt}}

\setcounter{thm}{0}
\renewcommand{\thethm}{S.\arabic{thm}}
\renewcommand{\thedfn}{S.\arabic{dfn}}
\renewcommand{\thelm}{S.\arabic{lm}}
\renewcommand{\thepro}{S.\arabic{pro}}

%\setcounter{figure}{0}
%\def\thefigure{S.\arabic{figure}}
%%%%%%%%%%%%%%%%%%%%%%%%%%%%%%%%%%%%%%%

%\makeatletter \renewcommand{\@biblabel}[1]{[S#1]} \makeatother

\onecolumngrid

%\begin{widetext}

\begin{center}
{\large \bf Supplemental Material for  \protect \\ 
  ``\nameoftitle'' }\\
\vspace*{0.3cm}
Naoto Shiraishi$^{1}$ and Mizuki Yamaguchi$^{1}$ \\
\vspace*{0.1cm}

{$^{1}${\it Graduate school of arts and sciences, The University of Tokyo} } 
\end{center}

\setcounter{equation}{0}
\renewcommand{\theequation}{S.\arabic{equation}}

In this Supplemental Material, we first present brief reviews of integrability and non-integrability, respectively.
These reviews will serve as a short summary of the research field of non-integrability, as well as a means to place our results properly in this context.
We then provide proofs of several lemmas used in the main text.

\tableofcontents

\section{Brief review of techniques and results on integrable systems}

\subsection{Yang-Baxter equation and the quantum inverse scattering method}

We first explain the Yang-Baxter equation (quantum inverse scattering method), which is a standard tool to solve the integrable systems and derive a family of infinite local conserved quantities.
A detailed explanation can be seen in e.g., a lecture note by Fadeev~\cite{s-Fad}.
Consider a one-parameter family of matrix $R_{ij}(\lmd)$ on two sites $i$ and $j$ with spin $S$ such that $R(0)$ is the swap operator $\Pi$, which is called {\it $R$ matrix}.
The Yang-Baxter equation is then expressed as
\eqa{
R_{12}(\lmd)R_{13}(\lmd+\mu)R_{23}(\mu)=R_{23}(\mu)R_{13}(\lmd+\mu)R_{12}(\lmd).
}{YB-s}

If there exists $R$ matrix satisfying the Yang-Baxter equation, then we can construct an integrable Hamiltonian by the following procedure.
We consider $N$ site $1,\ldots , N$ and one auxiliary site $a$ all of which have spin $S$.
Using $R$ matrix, we introduce the monodromy matrix $M(\lmd)$ on sites $1,2,\ldots ,N$ and $a$, and the transfer matrix $T(\lmd)$ on sites $1,2,\ldots ,N$ as
\balign{
M(\lmd )&:=\prod_{i=1}^N R_{ia}(\lmd), \\
T(\lmd )&:=\Tr _a[M(\lmd)].
}

Thanks to the Yang-Baxter equation, two transfer matrices commute
\balign{
[T(\lmd), T(\lmd')]&=0, \lb {prop-1}
}
which can be demonstrated as follows.
Since two operators acting on different sites commute, what we need to show is the middle equality of
\eq{
T(\lmd) T(\lmd')=\Tr_{a,b}[\prod_i R_{ia}(\lmd)R_{ib}(\lmd')]=\Tr_{a,b}[\prod_i R_{ib}(\lmd')R_{ia}(\lmd)]=T(\lmd')T(\lmd).
}
Plugging $I=R^{-1}_{ab}(\lmd-\lmd')R_{ab}(\lmd-\lmd')$ to the left end of the product and applying the Yang-Baxter equation \eqref{YB-s} repeatedly, we have
\balign{
&\Tr_{a,b}[\prod_i R_{ia}(\lmd)R_{ib}(\lmd')] \nt \\
=&\Tr_{a,b}[R^{-1}_{ab}(\lmd-\lmd')R_{ab}(\lmd-\lmd')R_{1a}(\lmd)R_{1b}(\lmd')R_{2a}(\lmd)R_{2b}(\lmd')\cdots R_{Na}(\lmd)R_{Nb}(\lmd')] \nt \\
=&\Tr_{a,b}[R^{-1}_{ab}(\lmd-\lmd')R_{1b}(\lmd')R_{1a}(\lmd)R_{ab}(\lmd-\lmd')R_{2a}(\lmd)R_{2b}(\lmd')\cdots R_{Na}(\lmd)R_{Nb}(\lmd')] \nt \\
=&\Tr_{a,b}[R^{-1}_{ab}(\lmd-\lmd')R_{1b}(\lmd')R_{1a}(\lmd)R_{2b}(\lmd')R_{2a}(\lmd)R_{ab}(\lmd-\lmd')\cdots R_{Na}(\lmd)R_{Nb}(\lmd')] \nt \\
=&\cdots \nt \\
=&\Tr_{a,b}[R^{-1}_{ab}(\lmd-\lmd')R_{1b}(\lmd')R_{1a}(\lmd)R_{2b}(\lmd')R_{2a}(\lmd)\cdots R_{Nb}(\lmd')R_{Na}(\lmd)R_{ab}(\lmd-\lmd')] \nt \\
=&\Tr_{a,b}[\prod_i R_{ib}(\lmd')R_{ia}(\lmd)],
}
which implies \eref{prop-1}.

Consider the Taylor series of $f(T(\lmd))$ for any function $f$ with respect to $\lmd$ expanded as
\eq{
f(T(\lmd))=\sum_{n=0}^\infty F_n \lmd^n.
}
Since \eref{prop-1} implies 
\eqa{
[f(T(\lmd), f(T(\mu))]=0,
}{ff-commute}
comparing the coefficient of $\lmd^n \mu^m$ in \eref{ff-commute} we find 
\eq{
[F_n, F_m]=0.
}

In particular, we set $f(x)=\ln x$ in the quantum inverse scattering method.
Then, $F_{n-1}$ serves as the $n$-local conserved quantity, which is computed simply by differentiating $\ln T(\lmd)$ as
\eqa{
Q^n=\frac{F_{n-1}}{(n-1)!}=\ft{\frac{d^{n-1}}{d\lmd ^{n-1}}\ln T(\lmd)}{\lmd=0}.
}{Qn}
By regarding $Q^2$ as the Hamiltonian $H$, the obtained $Q^n$ is the $n$-local conserved quantity of $H$.
The $n$-localness of $Q^n$ is confirmed by the boost operator argument presented in \sref{boost}

\subsection{Reshetikhin condition}

Consider an $R$ matrix satisfying the Yang-Baxter equation \eqref{YB-s}.
We can expand the $R$ matrix as
\eqa{
R(\lmd)=\Pi+\lmd \Pi h +\lmd^2 \Pi R^{(2)}+\lmd^3 \Pi R^{(3)}+\cdots,
}{R-expand-s}
where $h$ is the local Hamiltonian of $H$ and $\Pi$ is a swap operator.

For our purpose, it is useful to define 
\eq{
R'(\lmd)=\Pi R(\lmd),
}
which is expanded as
\eqa{
R'(\lmd)=I+\lmd h +\lmd^2 R^{(2)}+\lmd^3 R^{(3)}+\cdots
}{modR-expand}
and satisfies a modulated version of the Yang-Baxter equation
\eqa{
R'_{12}(\lmd)R'_{23}(\lmd+\mu)R'_{12}(\mu)=R'_{23}(\mu)R'_{12}(\lmd+\mu)R'_{23}(\lmd).
}{mod-YB}
We plug \eref{modR-expand} into the Yang-Baxter equation \eqref{mod-YB} and compare the coefficients of $\lmd$ and $\mu$.

The coefficients of $\lmd\mu$ in \eref{mod-YB} reads
\eq{
2R^{(2)}_{23}-h^2_{23}=2R^{(2)}_{12}-h^2_{12},
}
which implies
\eq{
R^{(2)}=\frac12 h^2.
}

The coefficients of $\lmd\mu^2$ reads
\eq{
R^{(3)}_{12}-R^{(3)}_{23}+\frac16 [h_{12}+h_{23}, [h_{12}, h_{23}]]-\frac16 h^3_{12}+\frac16 h^3_{23}=0.
}
Defining unknown 2-support operator $E:=h^3-6R^{(3)}$, we arrive at the Reshetikhin condition~\cite{s-KS82, s-Ken92}
\eqa{
D_{123}=-[h_{12}+h_{23}, [h_{12}, h_{23}]]=E_{23}-E_{12}.
}{Res-s}
The Reshetikhin condition is thus the lowest-order nontrivial relation of the Yang-Baxter equation.
This relation is sometimes employed to seek numerically new integrable models~\cite{s-LMS25}.

\subsection{Boost operator}\lb{s:boost}

We define the boost operator $B$ as
\eq{
B=\sum_j j h_{j,j+1}.
}
We do not care about the effect of boundaries since it is irrelevant to our analysis.

We first suppose that a Hamiltonian satisfies the Yang-Baxter equation \eqref{YB-s}.
Then, the corresponding transfer matrix $T(\lmd)$ satisfies
\eqa{
[B, T(\lmd)]=\frac{d}{d\lmd}T(\lmd),
}{BT-com}
which is shown as follows~\cite{s-Tel82}.

Differentiating the Yang-Baxter equation \eqref{YB-s} by $\mu$ and setting $\mu=0$ with multiplying $\Pi_{23}$ from right, we find
\eqa{
\( \frac{d}{d\lmd}R_{12}(\lmd)\) R_{13}(\lmd) -R_{12}(\lmd)\( \frac{d}{d\lmd}R_{13}(\lmd) \) =[ R_{12}(\lmd)R_{13}(\lmd), h_{2,3}],
}{Sutherland}
where we used $R(0)=\Pi$ and $\ft{\frac{d}{d\mu}R(\mu)}{\mu=0}=\Pi h$.
This relation is sometimes called the {\it Sutherland equation}.
We reassign subscripts as $1\to a$, $2\to i$, $3\to i+1$, and multiply $\prod_{j\leq i-1}R_{ja}(\lmd)$ from left and $\prod_{j\geq i+2}R_{ja}(\lmd)$ from right, which leads to
\eq{
\prod_{j\leq i-1}R_{ja}(\lmd)\( \frac{d}{d\lmd}R_{ia}(\lmd)\) \prod_{j\geq i+1}R_{ja}(\lmd) -\prod_{j\leq i}R_{ja}(\lmd)\( \frac{d}{d\lmd}R_{i+1,a}(\lmd)\) \prod_{j\geq i+2}R_{ja}(\lmd) =[ \prod_j R_{ja}(\lmd), h_{i,i+1}].
}
Finally, we multiply $-i$ and take the sum with respect to $i$, which reads
\eq{
\frac{d}{d\lmd} \prod_j R_{ja}(\lmd)=[B, \prod_j R_{ja}(\lmd)].
}
Here we used the telescope-type trick.
Ignoring the boundaries and taking its trace with $a$, we arrive at \eref{BT-com}.

Using \eref{BT-com}, we find
\eqa{
[B, T^m (\lmd)]=\sum_{l=0}^m T^l(\lmd)[B, T(\lmd)] T^{l-m-1}(\lmd)= \sum_{l=0}^m T^l(\lmd)\frac{d}{d\lmd}T(\lmd) T^{l-m-1}(\lmd) =\frac{d}{d\lmd}T^m(\lmd) ,
}{BTm-com}
and thus, considering the Taylor expansion we find
\eqa{
[B, f(T(\lmd))]=\frac{d}{d\lmd}f(T(\lmd)) 
}{BTf-com}
for any analytic function $f$.
Combining \eref{BTf-com} with \eref{Qn} by setting $f(x)=\ln x$, we find the recursive generation of local conserved quantities:
\eqa{
[B, Q^m]=\ft{\frac{d^m}{d\lmd^m} [B, \ln T(\lmd)]}{\lmd=0}=\ft{\frac{d^{m+1}}{d\lmd^{m+1}} \ln T(\lmd)}{\lmd=0} =Q^{m+1}.
}{B-recursive}
This relation guarantees that $Q^m$ is indeed an $m$-local quantity.

\bigskip

We next consider a general Hamiltonian $H$ on which the existence of the Yang-Baxter equation is not assumed.
In this case, the obtained $m$-support operator $Q^m$ by \eref{B-recursive} is no longer proven to be a conserved quantity at present.

The boost relation on the first order is
\eq{
Q^3=[B,H]=\sum_i [h_{i,i+1},h_{i+1,i+2}].
}
Whether this $Q^3$ conserves or not is open at present.
The conservation of $Q^3$ is expressed as
\eqa{
[[B,H],H]=0,
}{BHH}
which is also equivalent to the Reshetikhin condition \eqref{Res-s}.

If \eref{BHH} (or the Reshetikhin condition) holds, then $Q^4$ is also the 4-local conserved quantity, which is proven as
\eq{
[Q^4,H]=[[B, Q^3], H]=-[[Q^3,H],B]-[[H,B], Q^3]=0.
}
The conservation of $Q^m$ with $m\geq 5$ has recently been solved by Hokkyo:

\blm{[Hokkyo~\cite{s-Hok25-2}]\lb{t:Q3-Qm}
Consider a shift-invariant nearest-neighbor interaction Hamiltonian $H=\sum_i h_{i,i+1}$.
Suppose that $Q^3=[B,H]$ is a conserved quantity; $[Q^3,H]=0$.
Then, $Q^m$ given by \eref{B-recursive} are also conserved quantities; $[Q^m,H]=0$.
}

We here draw the outline of the proof of \lref{Q3-Qm}.
We prove $[Q^m,H]=0$ by induction on $m$.
We first show that $Q^m$ is shift-invariant.
Then, using the Jacobi identity we have
\eq{
[H, [H, Q^{m+1}]]=[H, [H, [B, Q^m]]]=[H, [Q^m, [B, H]]]=-[Q^m, [Q^3, H]]-[Q^3, [H, Q^m]]=0.
}
Finally, we notice that $[H, [H, Q^{m+1}]]=0$ implies $[H, Q^{m+1}]=0$, since the action of $[H, \cdot \ ]$ maps nonzero off-diagonal elements with respect to the eigenstates of $H$ onto nonzero ones (i.e., if $Q^{m+1}$ has a nonzero off-diagonal element, then the same off-diagonal element of $[H, Q^{m+1}]$ is still nonzero, implying $[H, [H, Q^{m+1}]]\neq 0$).
This completes the proof.

(We note that Zhang~\cite{s-Zha25} also claims the same statement as \lref{Q3-Qm}.
However, as the authors have examined, this proof has a logical gap.)

\section{Brief review of non-integrability proofs}\lb{s:review-nonint}

\subsection{Several definitions and proof idea}

We here briefly review several results and ideas of the proof of non-integrability.
We first introduce several key notions and then explain the proof ideas.
A detailed pedagogical introduction of non-integrability proofs can be found in Ref.~\cite{s-Shi24}.

Throughout this Letter and Supplemental Material, we define non-integrability for a Hamiltonian with nearest-neighbor interaction by the absence of nontrivial local conserved quantities.
To state it rigorously, we introduce the notion of a {\it $k$-support operator} such that its minimum contiguous support is among $k$ sites.
%In the case without confusion, the sum of $k$-support operators is also called simply a $k$-support operator.
For example, we state that $X_4Y_5Z_6$ is a 3-support operator and $X_2Y_5$ is a 4-support operator, since their contiguous supports are $\{4,5,6\}$ and $\{2,3,4,5\}$, respectively.
A conserved quantity is a $k$-local conserved quantity if this quantity is given by a sum of up to $k$-support operators and cannot be expressed by a sum of at most $k-1$-support operators.
The absence of nontrivial local conserved quantities in this case is defined by the absence of $k$-local conserved quantities for $k=O(1)$ with $k\geq 3$.
Here, the word ``nontrivial" means that the size of its contiguous support is strictly larger than that of the Hamiltonian.

The proof of non-integrability of spin chains is performed in the following procedure.
We first expand the operator space acting on a single spin $S$ with respect to a proper basis.
Then, we expand a candidate of a shift-invariant local conserved quantity up to $k$-support operators as
\eqa{
Q^k=\sum_{l=1}^{k} Q^{k,l}, \hspace{15pt} Q^{k,l}:=\sum_{\bsA^l} \sum_{i=1}^L q_{\bsA^l}\bsA^l_i
}{Qform}
with coefficients $q_{\bsA^l}\in \bbR$.
The sum of $\bsA^l$ runs over all possible sequences of basis operators with length $l$.
We notice that the commutator of a $k$-support operator $Q^k$ and $H$ is an at most $k+1$-support operator, which guarantees the expansion as
\eqa{
[Q^k,H]=\sum_{l=1}^{k+1}\sum_{\bsB^l} \sum_{i=1}^L r_{\bsB^l} \bsB^l_i.
}{QH}
Since the left-hand side can be expressed as a linear sum of $q_{\bsA}$ by inserting \eref{Qform}, and the conservation of $Q^k$ implies $r_{\bsB^l}=0$ for any $\bsB^l$, by comparing both sides of \eref{QH} we obtain many constraints (linear relations) on $q_{\bsA}$.
Our goal is to show that these linear relations do not have solutions except for $q_{\bsA ^k}=0$ for all $\bsA^k$, which means that $Q^k$ cannot be a $k$-local conserved quantity.

Concretely, we adopt the following two-step analysis.
\begin{quote}
\ul{Step 1}: We analyze linear relations in \eref{QH} with $l=k+1$ and restrict $k$-support operators which may have nonzero coefficients to a specific form.
In addition, we also show that the coefficients of the remaining $k$-support operators are linearly connected (i.e., $q_{\bsA}\propto q_{\bsA'}$).

\ul{Step 2}: We analyze linear relations in \eref{QH} with $l=k$ and demonstrate that one of the remaining coefficients of $k$-support operators is zero.
\end{quote}
It is observed that most non-integrability proofs employ essentially the same proof technique as above~\cite{Shi19, ST24, PL24-2, HYC24, YCS24-1, YCS24-2, FT25}, and Hokkyo~\cite{Hok25} demonstrates that this proof technique indeed works if two operator equations respectively admit solutions of specific types.

In the following two subsections, we first explain the proof technique for non-integrability by employing $S=1/2$ XYZ chain with $z$ magnetic field
\eq{
H=\sum_i h^{(2)}_{i,i+1}+h^{(1)}_i, \hspace{20pt} h^{(2)}_{i,i+1}=J_X X_iX_{i+1}+J_Y Y_iY_{i+1}+J_ZZ_iZ_{i+1}, \hspace{10pt} h^{(1)}_i=h_ZZ_i
}
as an example, and then briefly describe its general form for general non-integrability proofs.
Here, $X$, $Y$, and $Z$ are three Pauli operators.
We emphasize that the $S=1/2$ XYZ chain with $z$ magnetic field is employed just for the explanation of the proof technique, and this spin chain is not isotropic (i.e., it is not an example of our main theorem).
We also employ three Pauli matrices and the identity operator as the basis of operators ($2\times 2$ Hermitians), which is different from the noncommutative spherical harmonics $Y^{l,m}$.

\subsection{Step 1 of non-integrability proofs}

In step 1, we examine relations of $[Q^k,H]=\sum_{l=1}^{k+1}\sum_{\bsB^l} \sum_{i=1}^L r_{\bsB^l} \bsB^l_i=0$ with $l=k+1$.
To treat such relations transparently, we introduce a useful visualization of commutation relations.
For example, for $k=4$, 5-support operator $\sum_j X_jY_{j+1}X_{j+2}Y_{j+3}X_{j+4}$ in $[Q^4,H]$ is generated by the following two commutators:
\balign{
-i [X_jY_{j+1}X_{j+2}Z_{j+3}, Y_{j+3}Y_{j+4}]&=-2X_jY_{j+1}X_{j+2}X_{j+3}Y_{j+4}, \\
-i [Z_{j+1}X_{j+2}X_{j+3}Y_{j+4}, X_jX_{j+1}]&=2X_jY_{j+1}X_{j+2}X_{j+3}Y_{j+4},
}
where we dropped $\sum_j$, the summation over $j$, for visibility.
Using the {\it column expression}, we express that $X_jY_{j+1}X_{j+2}Y_{j+3}X_{j+4}$ is generated by these two commutation relations as follows:
\eq{
\begin{array}{rccccc}
&X&Y&X&Z& \\
&&&&Y&Y \\ \hline
&X&Y&X&X&Y
\end{array}
\ \ \ \ \
\begin{array}{rccccc}
&&Z&X&X&Y \\
&X&X&&& \\ \hline
&X&Y&X&X&Y
\end{array}, \nt %\lb{XYYX}
}
where we dropped all the coefficients in the outputs for brevity.
The horizontal positions represent the spatial positions of spin operators.
The column expression clearly shows that the operator $XYXXY$ in $[Q,H]$ is generated only by the above two commutators.
Recalling the condition $r_{XYXXY}=0$, we have the following relation
\eq{
-q_{XYXZ}+q_{ZXXY}=0.
}

Most of $k+1$-support operators are generated by two commutators.
On the other hand, some $k+1$-support operators are generated only by one commutator.
An example in $k=4$ is $YYYXZ$, which is generated only by
\eqa{
\begin{array}{rccccc}
&Y&Y&Y&Z& \\
&&&&Y&Y \\ \hline
&Y&Y&Y&X&Y
\end{array}.
}{step1-com-unique}
Here, the other commutator vanishes since no 4-support operator satisfies the following relation:
\eq{
\begin{array}{rccccc}
&&*&*&*&* \\
&Y&Y&&& \\ \hline
&Y&Y&Y&X&Y
\end{array}.
}
Then, \eref{step1-com-unique} leads to
\eq{
-J_Y q_{YYYZ}=0,
}
which directly implies $q_{YYYZ}=0$.

In addition, by considering commutators generating $XZYYZ$ as
\eq{
\ba{ccccc}{
&Y&Y&Y&Z \\
X&X&&& \\ \hline
X&Z&Y&Y&Z
}
\hspace{15pt}
\ba{ccccc}{
X&Z&Y&X& \\
&&&Z&Z \\ \hline
X&Z&Y&Y&Z
},
}
we readily have
\eq{
-J_Xq_{YYYZ}-J_Zq_{XZYX}=0.
}
However, since we have already shown $q_{YYYZ}=0$, we obtain $q_{XZYX}=0$.

Through similar arguments, we can show that a $k$-support operator has zero coefficient except for the case that this $k$-support operator can be constructed by repeatedly taking commutation relations with three interaction terms $XX$, $YY$, $ZZ$ with one-site shift as $[[[X_iX_{i+1}, Y_{i+1}Y_{i+2}], X_{i+2}X_{i+3}], Z_{i+3}Z_{i+4}]$.
From a more general perspective, such a $k$-support operator is an operator appearing in the {\it doubling-product operators} $\Di^k$ which is recursively defined as
\eq{
\Di^2:=h^{(2)}, \hspace{10pt} \Di^k_{1,2,\ldots ,k}:=[\Di^{k-1}_{1,2,\ldots , k-1}, h^{(2)}_{k-1,k}].
}
In addition, it is easy to confirm that all the remaining coefficients of $k$-support operators should be proportional to those of $\Di^k$.
In summary, we find
\eqa{
Q^{k,k}=c\Di ^k
}{step1-gen}
with a constant $c$.
This condition for $k=3$ means
\eqa{
Q^{3,3}=c \( J_XJ_Y XZY+J_YJ_Z YXZ +J_ZJ_X ZYX -J_YJ_X YZX-J_ZJ_Y ZXY-J_XJ_Z XYZ\) .
}{XYZz-step1}

This type of argument appears not only in the non-integrability proof for the $S=1/2$ XYZ chain with $z$ magnetic field but also in various non-integrability proofs~\cite{s-Shi19, s-Shi24, s-ST24, s-PL24-2, s-HYC24, s-YCS24-1, s-YCS24-2, s-Shi25, s-FT25}.
By a careful observation of these facts, Hokkyo~\cite{s-Hok25} demonstrated the wide applicability of this type of argument deriving \eref{step1-gen} if (i) $f(A)=[I_1\otimes A_2, h^{(2)}_{12}]$ and $g(A)=[A_1\otimes I_2, h^{(2)}_{12}]$ are injective, and (2) the operator equation for 2-support operator $X$
\eqa{
[X_{i,i+1}, h^{(2)}_{i+1,i+2}]+e^{2\pi i/l}[X_{i+1,i+2}, h^{(2)}_{i,i+1}]=0
}{step1-cond-s}
only has a solution $X\propto h^{(2)}$, where $l$ is a proper integer which divides the system size $L$.
The form of \eref{step1-gen} naturally arises since $k+1$-support operators in $[Q,H]$ on sites $1,2,\ldots , k+1$ come only from
\eq{
[Q^{k,k}_{1,\ldots , k}, h^{(2)}_{k,k+1}]+[Q^{k,k}_{2,\ldots , k+1}, h^{(2)}_{1,2}].
}
Imposing all $k+1$-support operators to have zero coefficients and comparing these two terms, we notice that $Q^{k,k}_{1,\ldots , k}$ should take the form of $[Y_{2,\ldots ,k}, {h^{(2)}}_{1,2}]$ with a $k-1$-support operator $Y$.
Applying similar arguments repeatedly, we obtain the doubling-product operator \eqref{step1-gen} if the last step (\eref{step1-cond-s}) holds.

\subsection{Step 2 of non-integrability proofs}

In step 2, we examine $k$-support operators in $[Q,H]$ and demonstrate that all the remaining coefficients of $k$-support operators in $Q$ (i.e., $Q^{k,k}$) to be zero.
Thanks to \eref{step1-cond-s} showing that all coefficients share the same factor $c$, it suffices to show the coefficient of one of the remaining $k$-support operators in $Q$ to be zero.

All the $k$-support operators in $[Q,H]$ on sites $1,2,\ldots ,k$ come only from
\eq{
[Q^{k,k-1}_{1,\ldots , k-1}, h^{(2)}_{k-1,k}]+[Q^{k,k-1}_{2,\ldots , k}, h^{(2)}_{1,2}]+\sum_{i=1}^{k-1} [Q^{k,k}, h^{(2)}_{i,i+1}] +\sum_{i=1}^{k} [Q^{k,k}, h^{(1)}_{i}] .
}
Note that this is a general expression and isotropic spin chains, which we treat in the main text, do not have the $h^{(1)}$ term.
In these linear relations, $Q^{k,k}$ is known, while $Q^{k,k-1}$ is unknown.
We extract several proper relations from these linear relations and solve them to show that one of the remaining coefficients of $k$-support operators to be zero.

We first treat the $S=1/2$ XYZ chain with $z$ magnetic field.
Although our result holds for general $k$, we here demonstrate the proof only for $k=3,4,5$ in order to see what happens clearly.
For $k=3$, we consider commutators generating $YYZ$ and $ZXX$ as
\eq{
\ba{ccc}{
X&Y&Z \\
Z&& \\ \hline
Y&Y&Z
}
\hspace{15pt}
\ba{ccc}{
Y&X&Z \\
&Z& \\ \hline
Y&Y&Z
}
\hspace{15pt}
\ba{ccc}{
Y&X& \\
&Z&Z \\ \hline
Y&Y&Z
}
}
and
\eq{
\ba{ccc}{
Z&X&Y \\
&&Z \\ \hline
Z&X&X
}
\hspace{15pt}
\ba{ccc}{
Z&Y&X \\
&Z& \\ \hline
Z&X&X
}
\hspace{15pt}
\ba{ccc}{
&Y&X \\
Z&Z& \\ \hline
Z&X&X
},
}
which lead to
\balign{
-h_Zq_{XYZ}-h_Zq_{YXZ}-J_{Z}q_{YX}&=0, \\
h_Zq_{ZXY}+h_Zq_{ZYX}+J_Zq_{YX}&=0.
}
Summing these two relations and plugging relations on $q_{\bsA}$ shown in \eref{XYZz-step1}, we find
\eq{
-2\( 1-\frac{J_Y}{J_X}\) h_Z q_{XYZ}=0.
}
This relation implies that $q_{XYZ}=0$ as long as $h_Z\neq 0$ (XYZ model) and $J_X\neq J_Y$ (XXZ model with $z$ magnetic field).

For $k=4$, we consider commutators generating $YYYX$, $ZXYZ$, and $XYXX$ as
\eq{
\ba{cccc}{
X&Y&Y&X \\
Z&&& \\ \hline
Y&Y&Y&X
}
\hspace{15pt}
\ba{cccc}{
Y&X&Y&X \\
&Z&& \\ \hline
Y&Y&Y&X
}
\hspace{15pt}
\ba{cccc}{
Y&Y&Z& \\
&&X&X \\ \hline
Y&Y&Y&X
}
}
and
\eq{
\ba{cccc}{
Z&Y&Y&Z \\
&Z&& \\ \hline
Z&X&Y&Z
}
\hspace{15pt}
\ba{cccc}{
Z&X&X&Z \\
&&Z& \\ \hline
Z&X&Y&Z
}
\hspace{15pt}
\ba{cccc}{
&Y&Y&Z \\
Z&Z&& \\ \hline
Z&X&Y&Z
}
\hspace{15pt}
\ba{cccc}{
Z&X&X& \\
&&Z&Z \\ \hline
Z&X&Y&Z
}
}
and
\eq{
\ba{cccc}{
X&Y&X&Y \\
&&&Z \\ \hline
X&Y&X&X
}
\hspace{15pt}
\ba{cccc}{
X&Y&Y&X \\
&&Z& \\ \hline
X&Y&X&X
}
\hspace{15pt}
\ba{cccc}{
&Z&X&X \\
X&X&& \\ \hline
X&Y&X&X
},
}
which lead to
\balign{
-h_Zq_{XYYX}-h_Zq_{YXYX}-J_Xq_{YYZ}&=0 \lb{step2-4-1} \\
h_Zq_{ZYYZ}-h_Zq_{ZXXZ}+J_Zq_{YYZ}-J_Zq_{ZXX}&=0 \lb{step2-4-2} \\
h_Zq_{XYXY}+h_Zq_{XYYX}+J_Xq_{ZXX}&=0. \lb{step2-4-3}
}
Summing \eref{step2-4-1}, \eref{step2-4-2} with multiplying $J_X/J_Z$, and \eref{step2-4-3}, and combining \eref{step1-gen} (which can be expanded as \eref{XYZz-step1}), we obtain
\eq{
-3h_Z \( 1-\frac{J_Y}{J_X}\) h_Z q_{XYYX}=0.
}
This relation implies that $q_{XYYX}=0$ as long as $h_Z\neq 0$ (XYZ model) and $J_X\neq J_Y$ (XXZ model with $z$ magnetic field).

For $k=5$, we consider commutators $YYYZY$, $ZXYYX$, $XYXYZ$, and $YZYXX$ as
\eq{
\ba{ccccc}{
X&Y&Y&Z&Y \\
Z&&&& \\ \hline
Y&Y&Y&Z&Y
}
\hspace{15pt}
\ba{ccccc}{
Y&X&Y&Z&Y \\
&Z&&& \\ \hline
Y&Y&Y&Z&Y
}
\hspace{15pt}
\ba{ccccc}{
Y&Y&Y&X& \\
&&&Y&Y \\ \hline
Y&Y&Y&Z&Y
}
}
and
\eq{
\ba{ccccc}{
Z&Y&Y&Y&X \\
&Z&&& \\ \hline
Z&X&Y&Y&X
}
\hspace{15pt}
\ba{ccccc}{
Z&X&X&Y&X \\
&&Z& \\ \hline
Z&X&Y&Y&X
}
\hspace{15pt}
\ba{ccccc}{
&Y&Y&Y&X \\
Z&Z&&& \\ \hline
Z&X&Y&Y&X
}
\hspace{15pt}
\ba{ccccc}{
Z&X&Y&Z& \\
&&&X&X \\ \hline
Z&X&Y&Y&X
}
}
and
\eq{
\ba{ccccc}{
X&Y&Y&Y&Z \\
&&Z&& \\ \hline
X&Y&X&Y&Z
}
\hspace{15pt}
\ba{ccccc}{
X&Y&X&X&Z \\
&&&Z& \\ \hline
X&Y&X&Y&Z
}
\hspace{15pt}
\ba{ccccc}{
&Z&X&Y&Z \\
X&X&&& \\ \hline
X&Y&X&Y&Z
}
\hspace{15pt}
\ba{ccccc}{
X&Y&X&X& \\
&&&Z&Z \\ \hline
X&Y&X&Y&Z
}
}
and
\eq{
\ba{ccccc}{
Y&Z&Y&Y&X \\
&&&Z& \\ \hline
Y&Z&Y&X&X
}
\hspace{15pt}
\ba{ccccc}{
Y&Z&Y&X&Y \\
&&&&Z \\ \hline
Y&Z&Y&X&X
}
\hspace{15pt}
\ba{ccccc}{
&X&Y&X&X \\
Y&Y&&& \\ \hline
Y&Z&Y&X&X
},
}
which lead to
\balign{
-h_Zq_{XYYZY}-h_Zq_{YXYZY}+J_Yq_{YYYX}&=0 \lb{step2-5-1} \\
h_Zq_{ZYYYX}-h_Zq_{ZXXYX}+J_Zq_{YYYX}+J_Xq_{ZXYZ}&=0 \lb{step2-5-2} \\
h_Zq_{XYYYZ}-h_Zq_{XYXXZ}+J_Xq_{ZXYZ}-J_Zq_{XYXX}&=0 \lb{step2-5-3} \\
h_Zq_{YZYYX}+h_Zq_{YZYXY}+J_Yq_{XYXX}&=0. \lb{step2-5-4}
}
Summing \eref{step2-5-1}, \eref{step2-5-2} with multiplying $-J_Y/J_Z$,  \eref{step2-5-3} with multiplying $J_Y/J_Z$, and \eref{step2-5-4}, and combining \eref{step1-gen}, we obtain
\eq{
-4h_Z \( 1-\frac{J_Y}{J_X}\) h_Z q_{XYYZY}=0.
}
This relation implies that $q_{XYYZY}=0$ as long as $h_Z\neq 0$ (XYZ model) and $J_X\neq J_Y$ (XXZ model with $z$ magnetic field).

As seen from the above $k=3$, 4, and 5 cases, the structure of step 2 is very similar among all $k$.
In fact, the arguments of general $k$ can be obtained by extending those for $k=3$, specifically by elongating the operators appearing in commutation relations for $k=3$ through the alternating insertion of $XX$ and $YY$, in order to construct $k$-support and $k-1$-support operators from 3-support and 2-support operators.
Hence, if we succeed in proving that one of the remaining 3-support operators has zero coefficients, then we can extend this argument to general $k$.

We encounter similar situation in various non-integrability proofs~\cite{s-Shi19, s-ST24, s-PL24-2, s-HYC24, s-YCS24-1, s-YCS24-2, s-FT25}.
Observing this situation, Hokkyo~\cite{s-Hok25} demonstrated that this elongation/reduction between general $k$-support operators and that for $k=3$ indeed works if (i) \eref{step1-gen} holds and (ii) we succeed in proving the absence of 3-local conserved quantity by step 2 (i.e., we need not examine 2-support and 1-support operators in $[Q,H]$).

\section{Derivation of injective condition}

Here we show that $f(A)=[I_1\otimes A_2, h_{12}]$ is injective for any nontrivial isotropic $h_{12}$.
Recalling the expression \eqref{YYform} and the fact that $\{ Y^{l,m}\} _{l,m}$ is a basis of the operator space, our claim is equivalent to the following lemma.

\blm{\lb{t:A-injective}
Fix the azimuthal quantum number $l>0$.
If $[A, Y^{l,m}]=0$ holds for all $-l\leq m\leq l$, then $A$ is proportional to the identity operator $I$.
}

We prove this lemma by two steps.
Below, we denote by $V^l={\rm span}\{ Y^{l,m}\}$ an operator subspace with azimuthal quantum number $l$.

\blm{\lb{t:l-decompose}
Fix the azimuthal quantum number $l>0$.
If $[A, Y^{l,m}]=0$ holds for all $-l\leq m\leq l$, then $[A^l, Y^{l,m}]=0$ holds for all $-l\leq m\leq l$, where $A^l$ is the projection of $A$ onto $V^l$.
}

We learned the following proof by Yuki Furukawa.

%https://www.math.toronto.edu/murnaghan/courses/mat445/ch6.pdf

\bpf{
We notice that the action of $g\in$ SU(2) given by $A\to U(g)AU^\dagger(g)$ ($U(g)$ is generated by $e^{iY^{1,\alpha} t}$ with $\alpha=0,\pm1$) keeps $V^l$ invariant, which is directly confirmed by the fact that the commutation relation $[Y^{l,m}, Y^{1,\alpha}]$ contains operators only in the form of $Y^{l,m'}$.
Combining the invariance of $V^l$ and our assumption, we obtain $[A, U(g)Y^{l,m}U^\dagger(g)]=[U^\dagger(g)AU(g), Y^{l,m}]=0$.
We multiply $\chi_l^*(g)$, the character of $l$, and integrate $g$ with the Haar measure on SU(2), we find the desired relation $[\int dg \chi_l^*(g)U^\dagger(g)AU(g), Y^{l,m}]=[A^l, Y^{l,m}]=0$.
Here, we used the character-based projection operator formula $P_l(A)=\int dg \chi_l^*(g)U^\dagger(g)AU(g)$, which comes from the orthogonality of characters, and the fact that $V^l$ is an irreducible representation.
}

\blm{\lb{t:Y-injective}
For any $l_1, l_2>0$ and $m_1$, there exists $m_2$ such that $[Y^{l_1,m_1}, Y^{l_2,m_2}]\neq 0$.
}

\bpf{
We set $m_2=-l_2$ for $l_1\geq 0$ and $m_2=l_2$ for $l_1<0$.
If $l_1+l_2-1\leq 2S$, we set $l=l_1+l_2-1$, and if $l_1+l_2-1>2S$ we set $l$ as one of $2S$ or $2S-1$ such that $l_1+l_2+l$ is odd.
With this choice, $m=m_1+m_2$ satisfies $\abs{m}\leq \abs{l_2}\leq l$.

Under the above choice, $[Y^{l_1,m_1}, Y^{l_2,m_2}]$ has $Y^{l,m}$ with a nonzero coefficient.
The non-vanishing of the coefficient (i.e., no accidental zero) is confirmed in \lref{CG-saturate}.
}

Combining these two lemmas, we find that $[A, Y^{l,m}]=0$  for all $-l\leq m\leq l$ implies $A\propto I$, which completes the proof of \lref{A-injective}.

\section{Derivation of Eq.(\ref{YYform})}

Here we shall show that an SU(2)-symmetric spin chain Hamiltonian $H=\sum_i h_{i,i+1}$ can be expressed as \eref{YYform}, which we recast below:
\eqa{
h_{i,i+1}=\sum_{l=0}^{2S} c_l \( \sum_{m=-l}^l Y^{l,m}_iY^{l,-m}_{i+1}\) .
}{YYform-s}
Note that the SU(2) symmetry means
\balign{
[h_{i,i+1}, S^0]&=0 \lb{SU2-0} \\
[h_{i,i+1}, S^{\pm 1}]&=0 \lb{SU2-1}
}
with
\balign{
S^m&:=\sum_{i=1}^N Y_i^{1,m} \hspace{10pt} (m=0,\pm1).
}
Below we start with a general expression of $h_{i,i+1}$ without any symmetry
\eqa{
h_{i,i+1}=\sum_{l,l'=0}^{2S}\sum_{m=-l}^l \sum_{m'=-l'}^{l'} c_{lml'm'} Y^{l,m}_iY^{l',m'}_{i+1}
}{YYform-gen}
and restrict its possible form to \eref{YYform-s} by imposing the SU(2) symmetry: \eqref{SU2-0} and \eqref{SU2-1}.

\blm{
The coefficient $c_{l,l',m,m'}$ in \eref{YYform-gen} for a spin chain with \eref{SU2-0} is zero if $m\neq -m'$.
}

\bpf{
Since the total magnetic quantum number is the eigenvalue of the commutator with $S^0$, i.e.,
\eq{
[Y_i^{l,m},S^0]=mY_i^{l,m},
}
we find
\eq{
[h_{i,i+1}, S^0]=\sum_{l,l',m,m'}c_{lml'm'}[Y_i^{l,m}Y_{i+1}^{l',m'},S^0]=\sum_{l,l',m,m'}c_{lml'm'}(m+m')Y_i^{l,m}Y_{i+1}^{l',m'}.
}
Comparing this relation with \eref{SU2-0}, we find $c_{l,l',m,m'}=0$ for $m\neq-m'$.
}

\blm{
The coefficient $c_{l,l',m,-m}$ in \eref{YYform-gen} for a spin chain with \eref{SU2-0} and \eref{SU2-1} is zero if $l\neq l'$.
}

\bpf{
Recall that the coefficient $K^{l_1,l_2,l}_{m_1,m_2}$ in \eref{YYK} (in the main text), which is proportional to the Clebsh-Gordon coefficient, is zero if the triangular condition $\abs{l_1-l_2}\leq l\leq l_1+l_2$ is not satisfied.
In addition, we also have $K^{l_1,l_2,l}_{m_1,m_2}=K^{l_2,l_1,l}_{m_2,m_1}$ if $l_1+l_2+l$ is even and $K^{l_1,l_2,l}_{m_1,m_2}=-K^{l_2,l_1,l}_{m_2,m_1}$ if $l_1+l_2+l$ is odd.
Setting $l_2=1$, we find
\eq{
[Y_i^{l,m},S^{\pm1}]=[Y_i^{l,m},Y_i^{1,\pm1}]=2K^{l,1,l}_{m,\pm1}Y_i^{l,m\pm1}.
}
The coefficient $K^{l,1,l}_{m,\pm1}$ is always nonzero for $\abs{m}\leq l$ and $\abs{m\pm 1}\leq l$ (for its proof, see \lref{CG-l1l}), and zero otherwise.
Using this relation, we find
\eqa{
[Y_i^{l,m}Y_{i+1}^{l',-m}, S^{\pm1}]=2K^{l,1,l}_{m,\pm1}Y_i^{l,m\pm1}Y_{i+1}^{l',-m}+2K^{l',1,l'}_{-m,\pm1}Y_i^{l,m}Y_{i+1}^{l',-m\pm1}.
}{YYform-pf-mid0}

We shall show that $c_{l,l',m,-m}=0$ for $l\neq l'$.
Without loss of generality, we assume $m\geq 0$ and $l<l'$.
Other cases are treated in similar manners.
We insert \eref{YYform-pf-mid0} for $S^{-1}$ into \eref{SU2-1} and compare the coefficients of $Y_i^{l,m}Y_{i+1}^{l',-m-1}$, which leads to a linear relation
\eqa{
2K^{l',1,l'}_{-m,-1}c_{l,m,l',-m}+2K^{l,1,l}_{m+1,-1}c_{l,m+1,l',-m-1}=0.
}{YYform-pf-mid1}
If $\abs{m+1}\leq l$ (and hence $\abs{-m-1}\leq l'$) is satisfied, then both coefficients $K$ are nonzero and we have
\eqa{
c_{l,m,l',-m}\propto c_{l,m+1,l',-m-1}.
}{YY-derive-proportional-pre}
Thus, replacing $m$ by $m+n$ ($n=1,2,\ldots $) and applying \eref{YYform-pf-mid1} repeatedly, we find
\eqa{
c_{l,m,l',-m}\propto c_{l,m+1,l',-m-1} \propto c_{l,m+2,l',-m-2}\propto \cdots \propto c_{l,l,l',-l}.
}{YY-derive-proportional}

On the other hand, at $c_{l,l,l',-l}$ we have $K^{l,1,l}_{l+1,-1}=0$ but $K^{l',1,l'}_{-l, -1}\neq 0$, which leads to
\eq{
2K^{l',1,l'}_{-m,-1}c_{l,l,l',-l}=0.
}
Plugging this relation into \eref{YY-derive-proportional}, we arrive at $c_{l,m,l',-m}=0$.
}

\blm{
If $c_{l,m,l,-m}$ is nonzero with some $m$, then $c_{l,m',l,-m'}$ is also nonzero. 
}

\bpf{
A Similar argument to \eref{YYform-pf-mid1} for $S^{+1}$ yields
\eq{
c_{l,m,l',-m}\propto c_{l,m-1,l',-m+1}.
}

Using this relation and \eref{YY-derive-proportional-pre}, we find
\eq{
c_{l,l,l,-l}\propto c_{l,l-1,l,-l+1}\propto c_{l,l-2,l,-l+2}\propto \cdots \propto c_{l,-l,l,l},
}
which directly implies the desired property.
}

We renormalize $Y^{l,m}$ such that these constants of proportion are one, which leads to
\eq{
c_{l,l,l,-l}=c_{l,l-1,l,-l+1}=c_{l,l-2,l,-l+2}= \cdots =c_{l,-l,l,l}.
}
By setting $c_l=c_{l,l,l,-l}$, we arrive at the desired expression \eqref{YYform-s}.

\section{Demonstration that Eq.(\ref{step1-cond}) has solution only $X\propto h$}

We here show that any solution $X$ to Eq.(\ref{step1-cond}) must be proportional to $h$.
To show this, we first denote by $L:=\{l|c_i\neq 0\}$ the set of nonzero coefficients $c_l$ in \eref{YYform}.
% and by $l_{\rm min}$ and $l_{\rm max} the minimum and maximum elements in $L$.
By assumption, $L\neq \emptyset$.
We then introduce several lemmas.
\blm{
The coefficient of $Y^{l,m}Y^{l',m'}$ in $X$ with $l\notin L$ is zero.
}
\bpf{
We expand $X$ as $X=\sum_{l,m}Y^{l,m}A^{l,m}$, where $A^{l,m}$ is generally not a single noncommutative spherical harmonics but their proper sum.
We suppose $A^{l,m}$ is nontrivial for $l\notin L$ and derive a contradiction.

\lref{A-injective} guarantees that for any nontrivial $A^{l,m}$ there exist $l^*\in L$ and $m^*$ satisfying $[A^{l,m}, Y^{l^*,m^*}]\neq 0$.
Then, an operator in the form of $Y^{l,m}_i[A^{l,m}, Y^{l^*,m^*}]_{i+1}Y^{l^*,-m^*}_{i+2}$ is generated by $[X_{i,i+1}, h_{i+1,i+2}]$ but is not generated by $[X_{i+1,i+2}, h_{i,i+1}]$ since $l\notin L$.
This means that $Y^{l,m}_i[A^{l,m}, Y^{l^*,m^*}]_{i+1}Y^{l^*,-m^*}_{i+2}$ does not vanish in the left-hand side of \eref{step1-cond}, which is contradiction.
}

\blm{\lb{t:step1-key}
The coefficient of $Y^{l,m}Y^{l',m'}$ in $X$ with $l, l'\in L$ has a zero coefficient if $l\neq l'$ or $m\neq -m'$.
}
\bpf{
In this and the following proofs, we use the column expression of commutation relations introduced in \sref{review-nonint}.
Note that unlike \sref{review-nonint} the last row in the column expression represents only one of the operators contained in this commutator, not all of them.

Consider commutators in $[X_{i,i+1}, h_{i+1,i+2}]$ and $[X_{i+1,i+2}, h_{i,i+1}]$ generating operators $Y^{l,m}_iY^{1,\pm1}_{i+1}Y^{l',m'\mp1}_{i+2}$.
The commutator $[X_{i,i+1}, h_{i+1,i+2}]$ contains
\eqa{
\ba{ccc}{
Y^{l,m}&Y^{l',m'}& \\
&Y^{l',-m'\pm1}&Y^{l',m'\mp1} \\ \hline
Y^{l,m}&Y^{1,\pm1}&Y^{l',m'\mp1}
}
}{step1-pf-mid1-1}
if $\abs{m'}\leq l'$, and the commutator $[X_{i+1,i+2}, h_{i,i+1}]$ contains
\eqa{
\ba{ccc}{
&Y^{l,m\pm1}&Y^{l',m'\mp1} \\ 
Y^{l,m}&Y^{l,-m}& \\ \hline
Y^{l,m}&Y^{1,\pm1}&Y^{l',m'\mp1}
}
}{step1-pf-mid1-2}
if $\abs{m\pm 1}\leq l$.
These two relations imply a linear relation between coefficients of $Y^{l,m}Y^{l',m'}$ and $Y^{l,m\pm 1}Y^{l',m'\mp 1}$ as
\eq{
K^{l',l',1}_{m',-m'\pm1}c_{l'}q_{l,m,l',m'}+K^{l,l,1}_{m\pm 1,-m}c_{l}q_{l,m\pm1,l',m'\mp1}=0.
}
Both coefficients $K^{l,l,1}_{m\pm1,-m}$ and $K^{l',l',1}_{m',-m'\pm1}$ are nonzero if $\abs{m'}\leq l'$ and $\abs{m\pm 1}\leq l$ (for its proof, see \lref{CG-ll1}), which implies
\eqa{
q_{l,m,l',m'}\propto q_{l,m\pm1,l',m'\mp1}.
}{step1-pf-mid2}
Applying \eref{step1-pf-mid2} repeatedly, we obtain a sequence
\eqa{
q_{l,m,l',m'}\propto q_{l,m\pm1,l',m'\mp1}\propto q_{l,m\pm2,l',m'\mp2}\propto q_{l,m\pm3,l',m'\mp3}\cdots .
}{step1-pf-propto}
This sequence finally reaches $\abs{m}=l$ or $\abs{m'}=l'$.

On the other hand, if exactly one of $\abs{m'}\leq l'$ or $\abs{m\pm 1}\leq l$ is not satisfied, then the remaining coefficient is shown to be zero.
In the case with $\abs{m'}\leq l'$ and $\abs{m\pm 1}> l$, we have $K^{l,l,1}_{m\pm 1,-m}=0$ and $K^{l',l',1}_{m',-m'\pm1}\neq 0$, which leads to
\eqa{
c_{l'}q_{l,m,l',m'}=0.
}{step1-pf-mid3-1}
In the case with $\abs{m'}>l'$ and $\abs{m\pm 1}\leq l$, we have $K^{l',l',1}_{m',-m'\pm1}=0$ and $K^{l,l,1}_{m\pm 1,-m}\neq 0$, which leads to
\eqa{
c_{l}q_{l,m\pm1,l',m'\mp1}=0.
}{step1-pf-mid3-2}
%In both cases, we readily find that the remaining coefficient ($q_{l,m,l',m'}$ or $q_{l,m\pm1,l',m'\mp1}$) to be zero.

Let us consider the case that one of $l=l'$ or $m=-m'$ is not satisfied.
Due to the symmetry of $(l,m)\lr (l',m')$ and $(m,m')\lr (-m,-m')$, without loss of generality we can assume $l-m<l'+m'$.
Other cases are treated in similar manners.
In the case of $l-m<l'+m'$, the $q_{l,m+n,l',m'-n}$ sequence in \eref{step1-pf-propto} reaches $q_{l,l,l',m'+m-l}$.
Noting $\abs{m'+m-l}\leq l'$, we apply \eqref{step1-pf-mid3-1} and find that all the coefficients in this sequence including $q_{l,m,l',m'}$ are zero.
}

\blm{
The coefficient of $Y^{l,m}Y^{l,-m}$ in $X$ is proportional to that in $h$.
}

\bpf{
We have shown in \lref{step1-key} that $X$ claims
\eq{
X=\sum_{l,m} d_{l,m}Y^{l,m}Y^{l,-m},
}
where we denote $d_{l,m}=q_{l,m,l,-m}$.
This lemma is equivalent to 
\eqa{
\frac{d_{l,m}}{d_{l',m'}}=\frac{c_l}{c_{l'}}
}{step1-fin-claim}
for any $l,m,l',m'$.

We first consider the case of $l=l'$.
We employ the same sequence obtained in the proof of \lref{step1-key}.
We set $l=l'$ and $m'=-m$ in Eqs.~\eqref{step1-pf-mid1-1} and \eqref{step1-pf-mid1-2}, which leads to
\eq{
K^{l,l,1}_{-m,m\pm1}c_l d_{l,m}+K^{l,l,1}_{m\pm1, -m}c_l d_{l,m\pm1}=0.
}
Substituting $K^{l,l,1}_{a,b}=-K^{l,l,1}_{b,a}$, which follows from oddness of $l+l+1$, into the above relation and replacing $m$ by $m\pm n$ repeatedly, we arrive at
\eq{
d_{l,m}=d_{l,m\pm 1}=d_{l,m\pm2}=\cdots ,
}
thereby confirming \eref{step1-fin-claim} in the case $l=l'$.
We denote the value of the above equation simply by $d_l$.

%To complete the proof of this lemma, it suffices to show that for any $l$ and $l'$, $d_{l,m}/d_{l',m'}=c_l/c_{l'}$ holds for some $m$ and $m'$.
We then treat the case of $l\neq l'$ and show $d_l/d_{l'}=c_l/c_{l'}$.
\lref{Y-injective} confirms that for any $l$ and $l'$ there exist $m$ and $m'$ such that $Y^{l,m}$ and $Y^{l',m'}$ do not commute.
Let $A$ be an operator which has a nonzero coefficient $K_a$ in $[Y^{l,m}, Y^{l',m'}]$.
Then, we consider commutators in $[X_{i,i+1}, h_{i+1,i+2}]$ and $[X_{i+1,i+2}, h_{i,i+1}]$ generating operators $Y^{l,-m}_iA_{i+1}Y^{l',-m'}_{i+2}$.
The commutator $[X_{i,i+1}, h_{i+1,i+2}]$ contains
\eq{
\ba{ccc}{
Y^{l,-m}&Y^{l,m}& \\
&Y^{l',m'}&Y^{l',-m'} \\ \hline
Y^{l,-m}&A&Y^{l',-m'}
}
}
and the commutator $[X_{i+1,i+2}, h_{i,i+1}]$ contains
\eq{
\ba{ccc}{
&Y^{l',m'}&Y^{l',-m'} \\ 
Y^{l,-m}&Y^{l,m}& \\ \hline
Y^{l,-m}&A&Y^{l',-m'}
},
}
which lead to
\eq{
K_a d_lc_{l'} -K_a d_{l'}c_l=0.
}
This directly implies the desired relation $d_l/d_{l'}=c_l/c_{l'}$, which completes the proof.
}

\section{Absence of accidental zeros in some Clebsch-Gordan coefficients}

In our analyses, we have taken commutation relations and derived nontrivial relations on coefficients.
To perform these analyses, the coefficients of commutation relations $K$, which is proportional to the Clebsh-Gordan coefficient, should be nonzero.
%Otherwise, we cannot extract useful information on coefficients.
Almost all Clebsh-Gordan coefficients are nonzero if the triangle inequality raised in the main text is satisfied.
However, it is known that there exist some {\it accidental zeros} in Clebsh-Gordan coefficients even if these two conditions are satisfied~\cite{s-RR84, s-HHR09, s-Donbook}.
Although it is highly implausible, for completeness we here prove that the Clebsh-Gordan coefficients employed in our analyses are indeed nonzero, i.e., we do not face accidental zeros in our argument.

\blm{\lb{t:CG-ll1}
The coefficient $K^{l_1,l_2,l}_{m_1,m_2}$ proportional to the Clebsch-Gordan coefficient  takes a nonzero value when $l_1=l_2$, $l=1$, and $m_1=-m_2\pm 1$.
}

\bpf{
Thanks to the general expression of the Clebsch-Gordan coefficients~\cite{s-Bohbook}, $K^{l_1,l_2,l}_{m_1,m_2}$  is proportional to
\eqa{
\sum_{k=A}^B \frac{(-1)^k}{k!(l_1+l_2-l-k)!(l_1-m_1-k)!(l_2+m_2-k)!(l-l_2+m_1+k)!(l-l_1-m_2+k)!},
}{CG-gen}
where $A$ and $B$ are set to 
\balign{
A&=\max (0, l_2-l-m_1, l_1-l+m_2), \\
B&=\min (l_1+l_2-l, l_1-m_1, l_2+m_2),
}
with which all the entries of factorials are nonnegative.
When $j=l_1=l_2$, $l=1$, and $m=m_1=-m_2\pm 1$, $A$ and $B$ are
\balign{
A&=\max (0, j-1-m, j-1-m\pm 1), \\
B&=\min (2j-1, j-m, j-m\pm1),
}
and thus $B=A$ in both cases of $\pm$.
This means that \eref{CG-gen} is a sum of a single nonzero summand, which is trivially nonzero.
}

\blm{\lb{t:CG-l1l}
The coefficient $K^{l_1,l_2,l}_{m_1,m_2}$ proportional to the Clebsch-Gordan coefficient takes a nonzero value when $l_1=l$, $l_2=1$, and $m_2=\pm 1$.
}

\bpf{
We again employ the expression \eqref{CG-gen}.
When $l_1=l$, $l_2=1$, and $m_2=\pm 1$, $A$ and $B$ are
\balign{
A&=\max (0,1-l-m_1, -m_2), \\
B&=\min (1, l-m_1, 1-m_2).
}
In the case of $m_2=1$, we have $A=B=0$, and in the case of $m_2=-1$, we have $A=B=1$.
This means that \eref{CG-gen} is a sum of a single nonzero summand, which is trivially nonzero.
}

\blm{\lb{t:CG-saturate}
The coefficient $K^{l_1,l_2,l}_{m_1,m_2}$ proportional to the Clebsch-Gordan coefficient takes a nonzero value when $\abs{m_2}=l_2$, $(l_1,l_2,l)$ satisfies the triangle relation, and $l_1+l_2+l$ is odd.
}

\bpf{
We again employ the expression \eqref{CG-gen}.

We first treat the case of $m_2=-l_2$.
In this case, $A$ and $B$ are
\balign{
A&=\max (0,l_2-l-m_1, l_1-l-m_2), \\
B&=\min (l_1+l_2-l, l_1-m_1,0),
}
which directly implies $A=B=0$.

We next treat the case of $m_2=l_2$.
In this case, $A$ and $B$ are
\balign{
A&=\max (0,l_2-l-m_1, l_1-l+l_2), \\
B&=\min (l_1+l_2-l, l_1-m_1,2l_2),
}
which directly implies $A=B=l_1+l_2-l$.

In both cases, \eref{CG-gen} is a sum of a single nonzero summand, which is trivially nonzero.
}

\end{document}